\documentclass{optica-article}
\journal{opticajournal}
\articletype{Research Article}

\usepackage{amsmath,bm}
\usepackage{siunitx,booktabs}
\usepackage{placeins}

\begin{document}

\title{Dark and bright solitons in a multimodal lithium tantalate microresonator}

\author{
Huizong Zhu\authormark{1,$\dagger$},
Yuyao Guo\authormark{1,$\dagger$},
Xiangfei Cheng\authormark{1},
Yiqun Wu\authormark{1},
and Juanjuan Lu\authormark{1,*}
}

\address{\authormark{1}School of Information Science and Technology, ShanghaiTech University, Shanghai 201210, China}
\address{\authormark{$\dagger$}Huizong Zhu and Yuyao Guo contributed equally to this work.}
\email{\authormark{*}lujj2@shanghaitech.edu.cn}

\begin{abstract*}
Bright and dark solitons require distinct dispersion and excitation conditions, complicating their integration within one microresonator. Spatial-mode engineering realizes wavelength-addressed dark- and bright-soliton microcombs in a multimodal lithium tantalate on insulator (LTOI) microring under fixed input polarization. The higher-order family assists dark-soliton excitation through local coupling to a normal-dispersion family and independently supports bright solitons at spectrally separated pump resonances through its anomalous dispersion. Neither polarization switching nor inter-resonator tuning is required. This dual role combines coupling-assisted excitation with independent soliton generation, providing a multimodal resonator architecture for multistate coherent sources.
\end{abstract*}

\section{Introduction}

Dissipative Kerr solitons in optical microresonators are sustained by the interplay of dispersion, Kerr nonlinearity, external driving and intracavity dissipation. These systems provide a platform for investigating nonlinear wave dynamics and developing on-chip coherent sources~\cite{Kippenberg2018,Diddams2020,Shen2020}. The temporal structure and optical power distribution of each soliton state determine its suitability for different photonic applications. Under anomalous group-velocity dispersion (GVD), bright solitons form localized, high-intensity pulses that support optical frequency synthesis, dual-comb spectroscopy and ultrafast ranging~\cite{Herr2014,Brasch2016,Spencer2018,Suh2016DualComb,Trocha2018Ranging}. Under normal GVD, dark solitons appear as localized intensity depressions on a continuous-wave background. Their formation can be described by switching waves connecting distinct intracavity steady states~\cite{Xue2015,ParraRivas2016,Yu2022continuum,Lv2025}. The high-duty-cycle waveforms and higher pump-to-comb conversion efficiency of dark-soliton microcombs favor multiwavelength carrier generation for coherent communications and photonic signal processing~\cite{Xue2017LPR,Fulop2018,Bai2023PPU}. Integrating both states in a single microresonator allows one light source to provide distinct temporal and spectral outputs. This requires accommodating their different background dispersions and excitation conditions within a shared resonator.

Previous studies have realized bright and dark microcombs by controlling pump configurations, coupled resonators and polarization modes. Dual pumping produces dark--bright soliton bound states through cross-phase modulation~\cite{Zhang2022DarkBright}. Electrothermally tuned coupled resonators exploit the Moir\'e speedup effect to reconfigure dispersion, allowing selection among bright microcombs, dark microcombs and Brillouin lasing~\cite{Ji2024}. Orthogonally polarized mode families provide distinct dispersion channels that support bright- and dark-soliton generation in a single microresonator~\cite{Xia2023}. These approaches modify the nonlinear excitation conditions through the cavity-mode structure or pump configuration, each with associated control requirements. Achieving multimodal soliton state selectivity with a simple drive configuration remains a challenge for the integration of compact on-chip light sources.

Transverse-mode families within a single microresonator offer spatial degrees of freedom for configuring distinct soliton excitation channels in a fixed device geometry. Modes within the same polarization class have distinct spatial field distributions and therefore respond differently to waveguide geometry. Cross-section design can consequently provide normal and anomalous GVD in different mode families over a common spectral range~\cite{Herr2014PRL,Yang2016,Yi2017}. Differences in free spectral range (FSR) also cause the relative resonance positions of the families to vary with optical frequency. When resonances approach in frequency and effective interfamily coupling is present, avoided mode crossings (AMXs) can produce localized resonance-frequency shifts. Outside these interaction regions, the two families can remain spectrally separated~\cite{Herr2014PRL,Yang2016,Yi2017}. Such localized perturbations have been used to assist microcomb generation under normal GVD, including dark-soliton excitation~\cite{Xue2015,Liu2014Optica,Song2026NormalGVD}. A higher-order family involved in local coupling can also support independent soliton generation, provided it has the anomalous dispersion and sufficiently low loss required for bright-soliton formation. This extends its role beyond auxiliary resonance control, turning a transverse-multimode cavity into a functionally multimodal soliton resonator.

Lithium tantalate on insulator (LTOI) has enabled low-loss, high-quality-factor microresonators and integrated Kerr microcombs, providing a basis for transverse-mode dispersion engineering and nonlinear excitation~\cite{He2025PR,Cai2024OL,Zhu2025LTA}. Bright-soliton generation under anomalous GVD, including octave-spanning soliton microcombs, further demonstrates the platform's capacity to sustain soliton formation and broadband nonlinear processes~\cite{Wang2024Nature,Cai2025PRJ,Zhu2025LTA}. Combining transverse-mode families with distinct background dispersions offers a route to extending the accessible soliton states while enabling their selection within the same LTOI device.

\begin{figure}[!t]
\centering
\includegraphics[width=\linewidth]{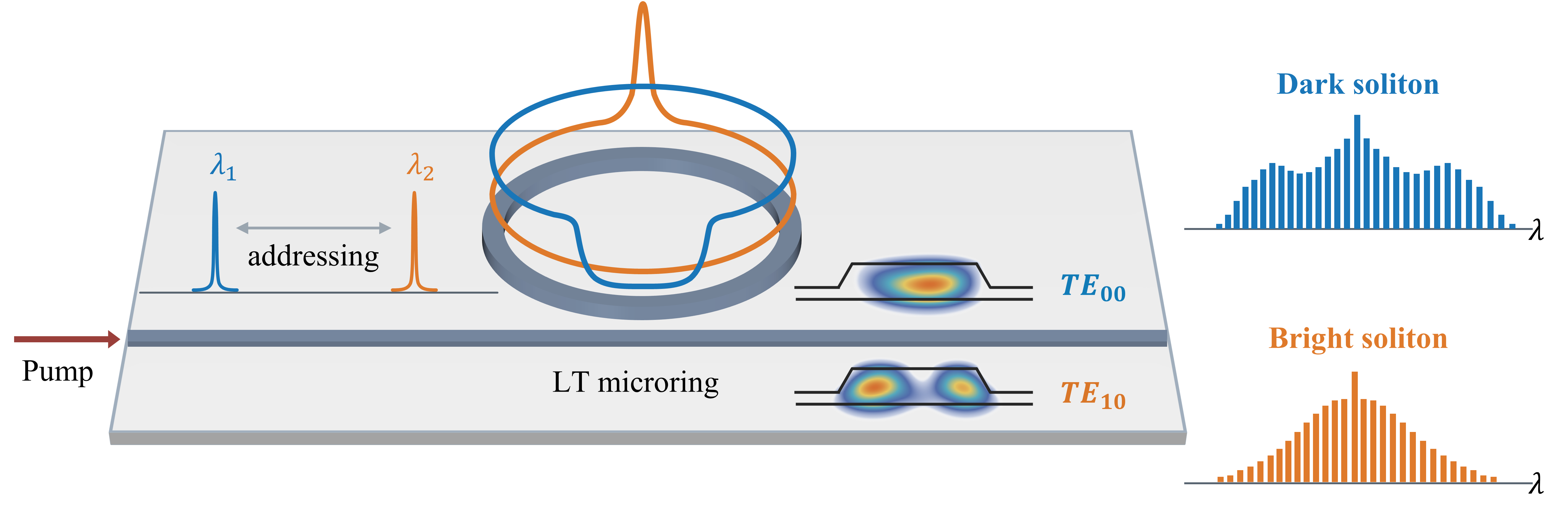}
\caption{
Wavelength-addressed dark- and bright-soliton microcombs in a LiTaO$_3$ microring.
A single tunable pump selectively addresses the TE$_{00}$ (blue) or TE$_{10}$ (orange) mode family at $\lambda_{1}$ or $\lambda_{2}$, respectively, with the input polarization held fixed.
The cross-sectional profiles show simulated electric-field intensities.
The intracavity waveforms and output spectra schematically illustrate dark- and bright-soliton operation in the normal- and anomalous-GVD families, respectively.
}
\label{fig:mode-family-selected-excitation}
\end{figure}

Here, spatial-mode engineering provides normal-GVD TE$_{00}$ and anomalous-GVD TE$_{10}$ transverse-electric (TE)-like mode families in a single LTOI microring, realizing wavelength-addressed dark- and bright-soliton microcombs under fixed input polarization [Fig.~\ref{fig:mode-family-selected-excitation}]. Mode-resolved resonance and dispersion measurements support the involvement of the higher-order TE$_{10}$ family in localized perturbations of the TE$_{00}$ resonance sequence that assist dark-soliton excitation. At spectrally separated pump resonances, the same family independently supports bright-soliton generation. Both microcombs are accessed by selecting the corresponding pump resonances, without polarization switching or inter-resonator tuning. To the best of our knowledge, this work presents the first experimental demonstration of a dark soliton microcomb in an integrated LTOI microring, extending the platform's soliton operation into the normal-GVD regime. By exploiting these two roles in different spectral regions of the same higher-order family, the design combines coupling-assisted excitation with independent soliton generation within a shared resonator. Spatial-mode engineering thus links the internal modal structure to wavelength-selective nonlinear excitation, providing a physical basis for integrating distinct nonlinear operating states within a multimodal photonic device.

\section{Device design and resonance characterization}
\begin{figure}[!htbp]
\centering
\includegraphics[width=\linewidth]{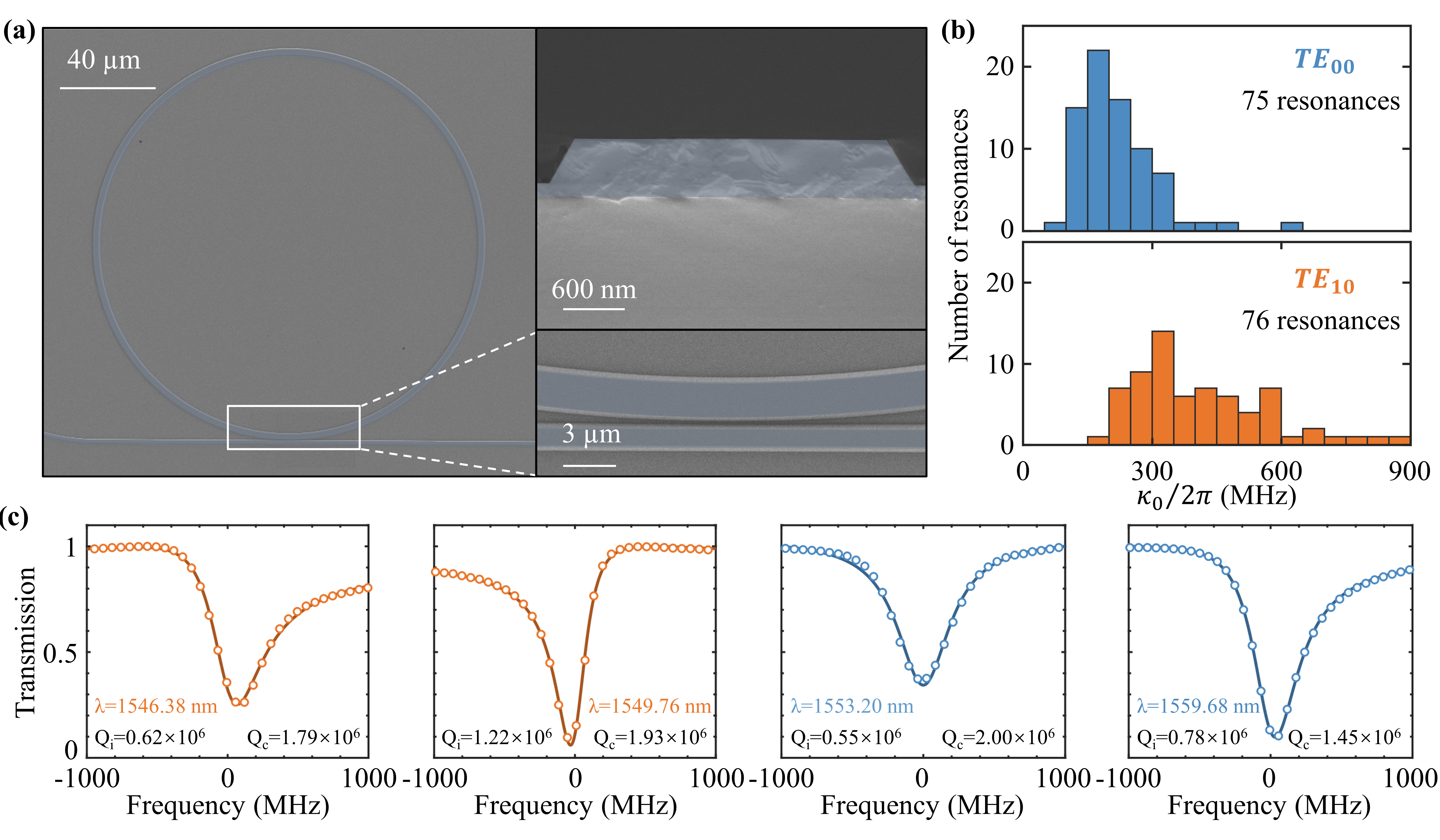}
\caption{
Device structure and resonance linewidths of the LiTaO$_3$ microring.
(a) Scanning electron microscopy (SEM) images of the complete microring (left), a cross-section of the $3~\mu\mathrm{m}$-wide input waveguide (upper right), and the coupling region between the bus waveguide and the microring (lower right).
LiTaO$_3$ waveguide regions are highlighted in false color. Scale bars, $40~\mu\mathrm{m}$, 600~nm and $3~\mu\mathrm{m}$, respectively.
(b) Intrinsic-linewidth distributions for 75 TE$_{00}$ and 76 TE$_{10}$ resonances.
Eight additional TE$_{10}$ resonances have intrinsic linewidths above 900~MHz and lie outside the displayed range.
(c) Transmission spectra of four resonances used for comb generation. Open circles denote measurements and solid curves denote fits.
From left to right, the resonances are TE$_{10}$ at 1546.38 and 1549.76~nm, followed by TE$_{00}$ at 1553.20 and 1559.68~nm.
Their intrinsic and coupling quality factors, $(Q_{i},Q_{c})$, are $(0.62,1.79)\times10^{6}$, $(1.22,1.93)\times10^{6}$, $(0.55,2.00)\times10^{6}$ and $(0.78,1.45)\times10^{6}$, respectively.
}
\label{fig:device-characterization}
\end{figure}

The microring cross-section was designed to support the TE$_{00}$ and TE$_{10}$ transverse-mode families with normal and anomalous background GVD, respectively. Numerical dispersion maps at 1560~nm guided the geometry selection (Supplement~1, Section~2 and Fig.~S1). We fabricated partially etched microrings on \textit{z}-cut LTOI using electron-beam lithography and Ar$^{+}$-based dry etching~\cite{Zhu2025LTA} (Supplement~1, Section~1). The device studied here has a nominal waveguide top width of $2.1~\mu\mathrm{m}$ and a radius of $80~\mu\mathrm{m}$ [Fig.~\ref{fig:device-characterization}(a)]. The LiTaO$_3$ film is 600~nm thick, with a 200-nm residual slab.

We measured the intrinsic linewidths of 75 TE$_{00}$ and 76 TE$_{10}$ resonances across the characterized spectral range [Fig.~\ref{fig:device-characterization}(b)]. Most TE$_{00}$ resonances have intrinsic linewidths below 300~MHz, whereas the TE$_{10}$ distribution is broader. At the four resonances selected for comb generation, the intrinsic quality factors are $(0.55$--$0.78)\times10^{6}$ for TE$_{00}$ and $(0.62$--$1.22)\times10^{6}$ for TE$_{10}$ [Fig.~\ref{fig:device-characterization}(c)]. The selected TE$_{10}$ resonances therefore retain intrinsic quality factors on the order of $10^{6}$, enabling direct pumping of the higher-order family for comb generation.

\section{Mode-family-resolved dispersion and local interactions}

\begin{figure}[tbp]
\centering
\includegraphics[width=\linewidth]{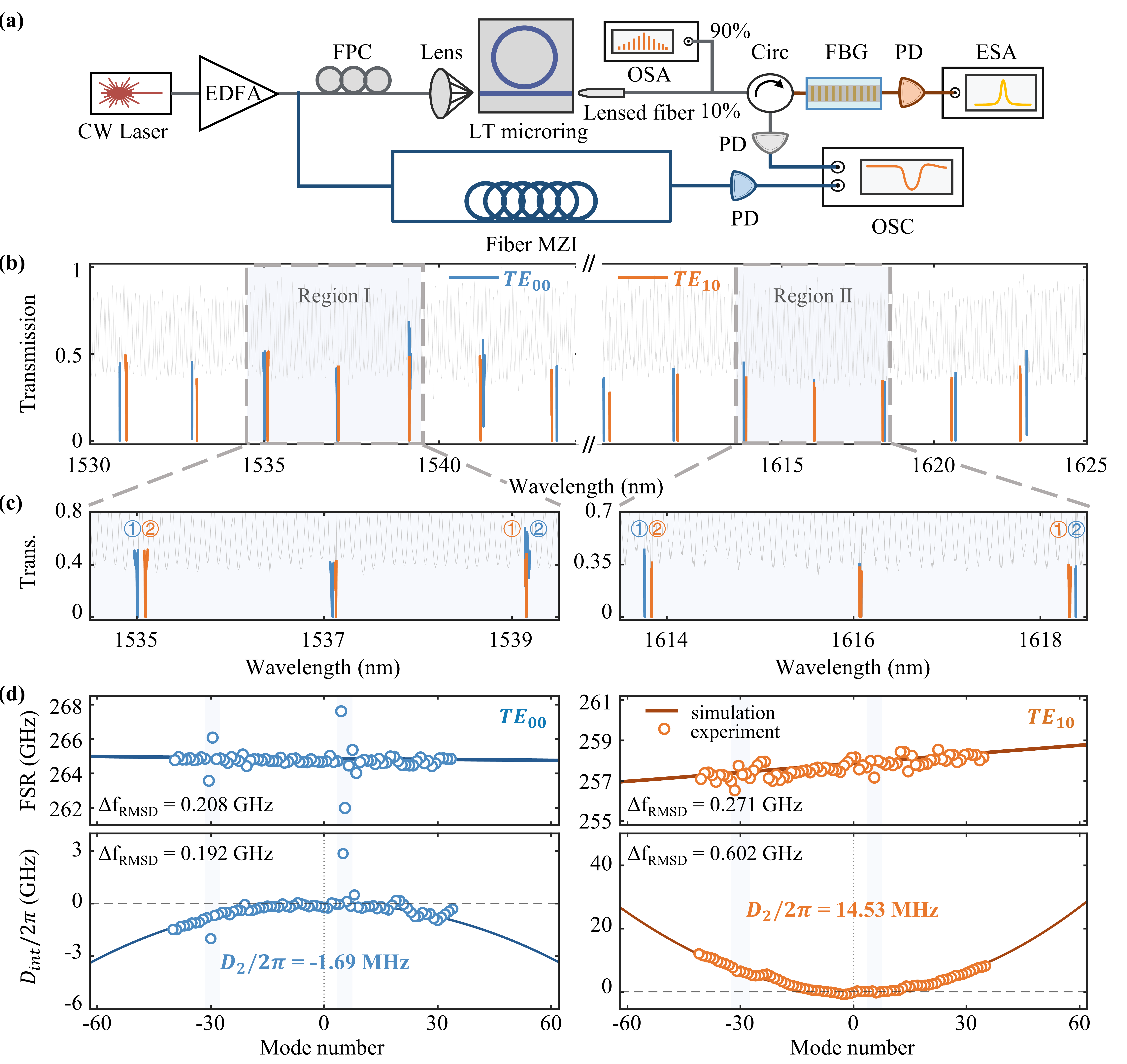}
\caption{
Mode-family-resolved spectroscopy, dispersion and local interactions.
(a) Experimental setup for dispersion characterization and microcomb generation.
The common pump/device path is combined with optical-spectrum readout, pump-suppressed RF detection and synchronous device-transmission/MZI-reference acquisition.
CW, continuous wave; RF, radio frequency; EDFA, erbium-doped fiber amplifier; FPC, fiber polarization controller; lens, free-space focusing lens; OSA, optical spectrum analyzer; Circ, optical circulator; FBG, fiber Bragg grating; PD, photodetector; ESA, electrical spectrum analyzer; MZI, Mach--Zehnder interferometer; OSC, oscilloscope.
(b) Two windows of the calibrated transmission spectrum.
The gray trace shows measured transmission; blue and orange markers identify the TE$_{00}$ and TE$_{10}$ resonance sequences.
Shading marks Regions~I and II, and the axis break separates noncontiguous wavelength intervals.
(c) Enlarged spectra within the shaded regions.
Numbered labels indicate the assigned resonances and changes in their relative spectral order.
(d) Local FSR and integrated dispersion $D_{\mathrm{int}}/2\pi$ for the two mode families.
Circles denote measurements and solid curves denote simulations.
Annotated root-mean-square deviations (RMSDs) quantify measurement--simulation differences, as defined in Supplement~1, Section~4 and Fig.~S3.
}
\label{fig:dispersion-characterization}
\end{figure}

We resolved the two mode families using broadband transmission spectroscopy with a tunable continuous-wave laser [Fig.~\ref{fig:dispersion-characterization}(a)]. A fiber Mach--Zehnder interferometer (MZI) calibrated the frequency axis over 1480--1640~nm, allowing both resonance sequences to be tracked across the measured band (Supplement~1, Fig.~S2). Resonance dips were fitted with Lorentzian profiles to extract their center frequencies. Comparison with simulated FSR and integrated-dispersion trends supports assigning the blue and orange sequences to TE$_{00}$ and TE$_{10}$, respectively. The measurement and analysis procedures, including the candidate-mode comparison, are provided in Supplement~1, Section~4 and Fig.~S3.

Each resonance sequence was indexed by a relative mode number $\mu$ that increases with optical frequency, with $\mu=0$ assigned near 1550~nm. The resonance frequencies are expanded as
\begin{equation}
\omega_{\mu}=\omega_{0}+D_{1}\mu+\frac{1}{2}D_{2}\mu^{2}+\frac{1}{6}D_{3}\mu^{3}+\cdots,
\label{eq:mode-expansion}
\end{equation}
where $D_{1}/2\pi$ is the first-order mode-spacing coefficient at the reference resonance [Eq.~\eqref{eq:mode-expansion}]. The local FSR and integrated dispersion are defined as
\begin{equation}
\mathrm{FSR}(\mu)=\frac{\omega_{\mu+1}-\omega_{\mu}}{2\pi},
\label{eq:local-fsr}
\end{equation}
and
\begin{equation}
D_{\mathrm{int}}(\mu)=\omega_{\mu}-(\omega_{0}+D_{1}\mu)
=\sum_{n\geq2}\frac{D_{n}}{n!}\mu^{n}.
\label{eq:dint}
\end{equation}
The relation $D_{2}\simeq-(c/n_{\mathrm{g}})D_{1}^{2}\beta_{2}$ connects the resonator dispersion to the waveguide GVD~\cite{Herr2014PRL}. Thus, $D_{2}<0$ and $D_{2}>0$ correspond to normal and anomalous GVD, respectively.

The mean FSRs are approximately 265~GHz for TE$_{00}$ and 258~GHz for TE$_{10}$. Outside the locally perturbed intervals, polynomial fits to the integrated-dispersion backgrounds yield $D_{2}/2\pi=-1.69$~MHz and $+14.53$~MHz, respectively [Fig.~\ref{fig:dispersion-characterization}(d)]. These opposite signs establish the normal- and anomalous-GVD channels within the same microring. The background fitting and residual analysis are described in Supplement~1, Section~4.

The approximately 7.0-GHz FSR mismatch causes the two resonance sequences to approach one another at selected wavelengths and separate elsewhere. It predicts a characteristic encounter interval of $N_{\mathrm{enc}}\sim\mathrm{FSR}/|\Delta\mathrm{FSR}|\approx37$ longitudinal orders, or approximately 79~nm near 1550~nm. The measured interaction regions, Region~I (1534.5--1539.5~nm) and Region~II (1613.5--1618.5~nm), have centers near 1537 and 1616~nm, consistent with this estimate [Fig.~\ref{fig:dispersion-characterization}(b)].

Within these regions, the TE$_{00}$ and TE$_{10}$ resonances approach one another and exchange their relative spectral order [Fig.~\ref{fig:dispersion-characterization}(c)]. Concurrent deviations in the local FSR and integrated dispersion support the participation of TE$_{10}$ in locally perturbing the TE$_{00}$ resonance sequence [Fig.~\ref{fig:dispersion-characterization}(d); Supplement~1, Fig.~S3]. These localized perturbations provide a mechanism for mode-interaction-assisted dark-soliton excitation on the normal-GVD background~\cite{Xue2015,Liu2014Optica}.

At the four selected pump resonances, the nearest resonances of the other family lie 31.6--72.9~GHz away. These separations correspond to approximately 75--192 loaded linewidths of the respective pumped resonances (Supplement~1, Section~3). The two families therefore interact locally while remaining spectrally separated at the selected pump resonances, providing the basis for wavelength-addressed excitation.

\FloatBarrier
\begin{figure}[!htbp]
\centering
\includegraphics[width=\linewidth]{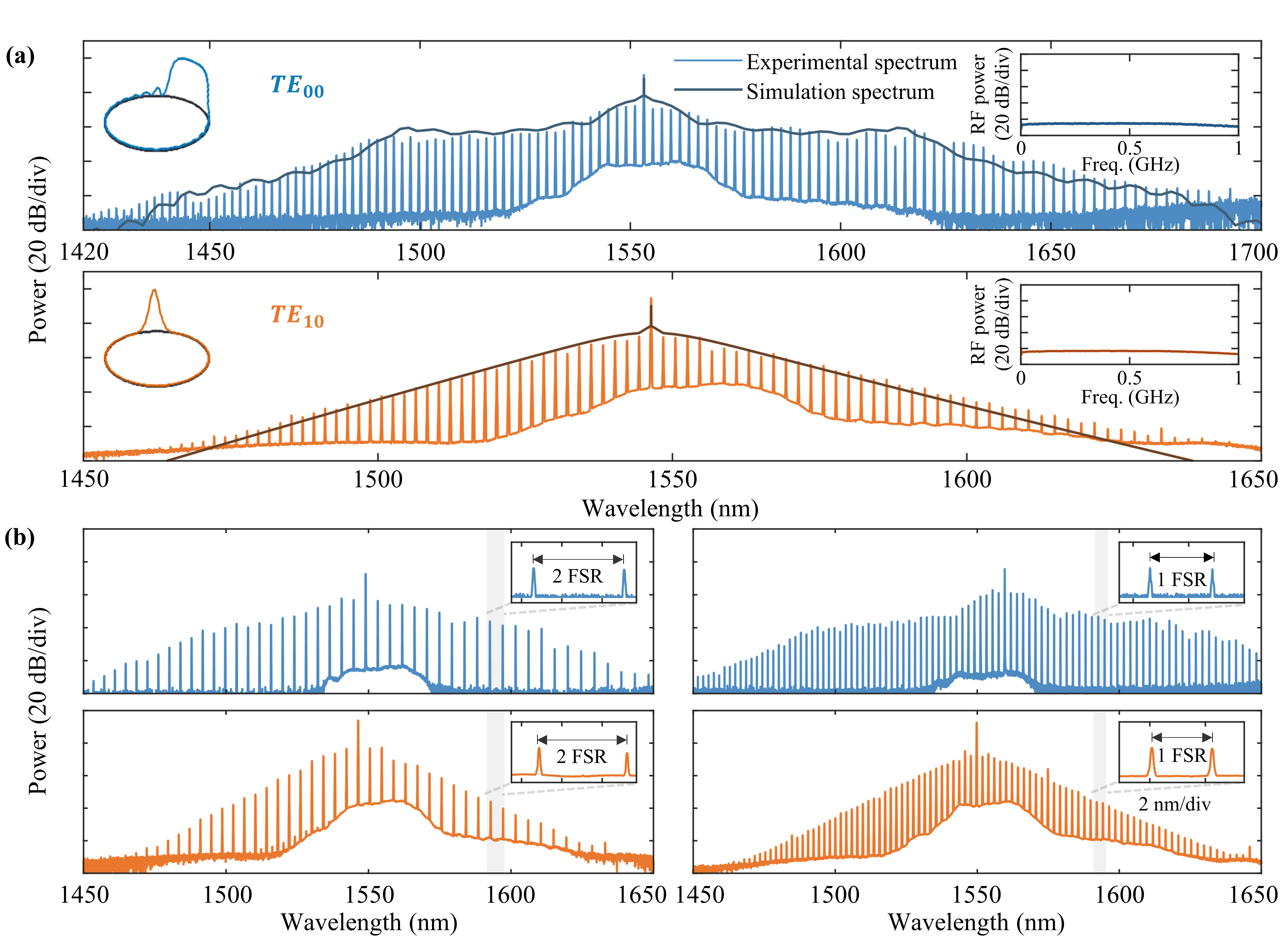}
\caption{
Wavelength-addressed dark- and bright-soliton microcombs under fixed input polarization.
(a) Measured spectra obtained by pumping TE$_{00}$ at 1553.20~nm (upper panel) and TE$_{10}$ at 1546.38~nm (lower panel).
Darker curves show the corresponding single-family Lugiato--Lefever equation (LLE) simulations.
Left insets show simulated intracavity intensity waveforms over one round trip in a circular representation; right insets show measured pump-suppressed baseband RF power spectra.
(b) Additional microcomb states obtained under TE$_{00}$ (upper panels) and TE$_{10}$ (lower panels) pumping, with spacings of two FSRs (left) and one FSR (right) of the respective mode families.
Insets resolve the local comb-line spacings.
The TE$_{10}$ spectra at 1546.38~nm in (a) and (b) correspond to different operating conditions at the same resonance.
}
\label{fig:comb-state-comparison}
\end{figure}
\section{Wavelength-addressed dark- and bright-soliton microcombs}

With the input polarization held fixed, the two microcomb states were accessed by pumping resonances of the corresponding mode families [Fig.~\ref{fig:comb-state-comparison}(a)]. The pump was slowly scanned from the red-detuned side towards each selected cavity resonance; the generation and detection procedures are detailed in Supplement~1, Section~5. Pumping TE$_{00}$ at 1553.20~nm produced the dark-soliton microcomb, whereas pumping TE$_{10}$ at 1546.38~nm produced the bright-soliton microcomb.

The two comb states were simulated separately using driven-dissipative single-family Lugiato--Lefever equations (LLEs)~\cite{Kippenberg2018,Herr2014,Lugiato1987,Chembo2013,Coen2013}. For each family, the background dispersion was approximated as $D_{\mathrm{int}}(\mu)=D_{2}\mu^{2}/2$ using the measured $D_{2}$, with local resonance shifts represented separately~\cite{Xue2015,Lobanov2015}. The equations, normalized formulation and numerical parameters are provided in Supplement~1, Section~6 and Table~S1.

The TE$_{00}$ dark-soliton microcomb spans approximately 1420--1690~nm. A single-family Lugiato--Lefever equation (LLE) simulation incorporating the measured normal GVD and localized resonance shifts reproduces its broadband spectral envelope. The simulated intracavity waveform contains high- and low-intensity regions connected by switching fronts, forming a dark-soliton structure over one cavity round trip [Fig.~\ref{fig:comb-state-comparison}(a), upper-left inset]. The localized resonance shifts represent the perturbations introduced by mode interactions in the normal-GVD channel.

By contrast, the TE$_{10}$ bright-soliton microcomb spans approximately 1480--1620~nm and exhibits a smooth $\mathrm{sech}^{2}$-like envelope. The corresponding LLE simulation, based on the measured anomalous GVD, reproduces this envelope and yields a bright-pulse waveform with a localized intensity peak [Fig.~\ref{fig:comb-state-comparison}(a), lower-left inset]. This result establishes the higher-order TE$_{10}$ family as a directly pumped bright-soliton channel, in addition to its participation in local mode interactions.

After pump suppression, neither microcomb exhibits pronounced noise peaks in the measured baseband radio-frequency (RF) power spectrum over 0--1~GHz [Fig.~\ref{fig:comb-state-comparison}(a), right insets]. These observations are consistent with low-noise operation over the measured frequency range. The simulated formation dynamics of the two states are presented in Supplement~1, Sections~7 and 8, respectively.

Both mode families also support microcomb states with different comb-line spacings, with representative spectra shown in Fig.~\ref{fig:comb-state-comparison}(b). Combs with spacings of one and two FSRs of the respective mode families were obtained in both the TE$_{00}$ dark-soliton and TE$_{10}$ bright-soliton channels. Both types of microcomb were generated at multiple pump resonances, while different states were also accessed at the same resonance. For example, the TE$_{10}$ spectra in Fig.~\ref{fig:comb-state-comparison}(a,b) were obtained by pumping the resonance at 1546.38~nm under different operating conditions. Thus, the dual-mode-family architecture supports wavelength-addressed dark- and bright-soliton generation, with different comb-line spacings accessible in both channels.

\section{Conclusion}

Spatial-mode engineering has enabled wavelength-addressed dark- and bright-soliton microcombs in a multimodal LTOI microring under fixed input polarization. Normal-GVD TE$_{00}$ and anomalous-GVD TE$_{10}$ mode families provide distinct excitation channels selected by the pump wavelength, without polarization switching or inter-resonator tuning. Both families support microcomb states with different comb-line spacings at multiple pump resonances. The dark-soliton demonstration extends LTOI microcomb operation into the normal-GVD regime.

The higher-order TE$_{10}$ family participates in localized TE$_{00}$ resonance perturbations that assist dark-soliton excitation, while supporting independent bright-soliton generation at spectrally separated pump resonances. This dual role combines coupling-assisted excitation with independent soliton formation in different spectral regions of the same mode family. The underlying combination of transverse-mode dispersion and localized interfamily coupling could be extended to other Kerr platforms, including lithium niobate and silicon nitride~\cite{Lv2025,Xue2015}. More broadly, these results bridge transverse spatial-mode engineering and functional multimodality in nonlinear microresonators. Harnessing internal modal degrees of freedom to unite coupling assisted and independent soliton regimes provides a generalizable strategy for compact, multimodal coherent light sources.

\begin{backmatter}

\bmsection{Funding}
This work was supported by the National Key R\&D Program of China (2024YFB2807400), the National Natural Science Foundation of China (62305214, 92576113), and the Shanghai Science and Technology Committee (25JD1405800).

\bmsection{Acknowledgment}
The facilities used for device fabrication were supported by the ShanghaiTech Material Device Lab (SMDL). J. Lu acknowledges support from the State Key Laboratory of Photonics and Communications, Shanghai Jiao Tong University, China.


\bmsection{Disclosures}
The authors declare no conflicts of interest..

\bmsection{Data availability}
Data underlying the results presented in this paper are not publicly available at this time but may be obtained from the authors upon reasonable request.

\bmsection{Supplemental document}
See Supplement 1 for supporting content.

\end{backmatter}

\bibliography{references_main}


\end{document}